\documentclass{PoS}

\def\ran{\rangle}

\title{The identification of glueballs - further tests}

\ShortTitle{Identification of glueballs}

\author{\speaker{Wolfgang Ochs}\\ 
        Max-Planck-Institut fuer Physik, Foehringer Ring 6, D-80805 Munich,
 Germany\\
        E-mail: \email{ochs@mpp.mpg.de}}

\abstract{The indirect evidence for gluonic objects in gluon jets is
recalled, more accurate tests at the LHC 
are possible.  Estimates of gluonic and quarkonic
  components of scalar mesons are presented, which could be 
 improved by further studies, in particular of $B$ and $B_s$ decays.
The results are consistent with the scalar glueball mixing into
$f_0(500)$ and $f_0(1500)$. The study of symmetry relations for 2-body
decay rates of charmonium $\chi_c$ is suggested as crucial test in the identification
  of $q\bar q$ multiplets and glueballs. Further related results are
discussed in the recent review \cite{Ochs:2013gi}.
}

\FullConference{XV International Conference on Hadron Spectroscopy-Hadron 2013\\
		4-8 November 2013\\
		Nara, Japan }

\begin{document}

\section{QCD expectations for glueballs} 

The identification of glueballs
remains a challenge in meson spectroscopy 
ever since they have been discussed within QCD as bound
states of the self-interacting gluons \cite{Fritzsch:1975tx}.  Today, on the
theory side, lattice QCD aims at predictions from basic principles. 
There are converging results within the theory of gluons only (``quenched'')
of a spectrum of $gg$ bound states, ``gluonium'', with the lightest state
being a scalar $J^{PC}=0^{++}$ of mass $1500-1700$ MeV.  In full QCD
the gluonium state can mix with $q\bar q$ states.  For the lightest scalar
flavour singlet state there are conflicting results: most recently the full
(``unquenched'') calculation yields comparable results to the purely gluonic
calculation \cite{Richards:2010ck} while earlier studies suggested 
the lightest states are of mass around 1000 and 1500 MeV and they are 
mixed from $q\bar q$ and $gg$ components \cite{Hart:2006ps}. 

An alternative path to spectroscopy is based on QCD sum rules which include
some phenomenological parameters. Already long
ago the lightest gluonium states have been located near 1000 MeV and 1500 MeV 
\cite{Narison:1988ts}. A more recent analysis finds that the states
at these two masses should be viewed as mixtures of $q\bar q$ and $gg$
components \cite{Harnett:2008cw}. 

A commonly accepted view of these QCD results  is still missing.
  
\section{Indications of glueball  production in gluon jets}

\begin{figure}[t]
\begin{center}
\includegraphics*[width=6.34cm,viewport=0.0cm 0.0cm 20.0cm %
20.0cm]{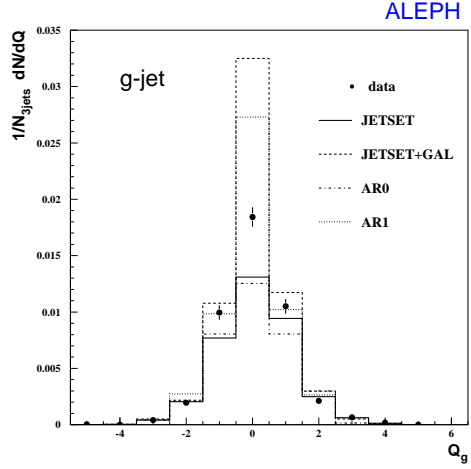}
\end{center}
\caption{Distribution of the charge $Q_g$ of the leading cluster 
in a gluon jet beyond a
rapidity gap $\Delta y=1.5$ showing a surplus of events at $Q_g=0$ 
over the
JETSET Monte Carlo prediction (full line) by $~40\%$.}
\label{fig:Q-Aleph}
\end{figure}

While there is a clear evidence for glueballs in the theory of gluons,
the evidence on the experimental side is still controversial; in particular,
there are very different interpretations of the scalar meson spectrum.
However, there is a remarkable effect which strongly points towards
the existence of gluonic objects. 
In a well known phenomenology of ``quark fragmentation'' 
the most energetic hadrons are those which carry a valence quark in common
with the initial quark of the jet, for example, the $\pi^+$ in a $u$-quark
jet. In analogy one can hypothesize that
glueballs are the leading objects in a gluon jet, an idea with a long
history \cite{Roy:1978pz}. As a practical tool one may isolate leading
clusters in jets by a rapidity gap: while in quark jets the charge of the
leading $q\bar q$ object is $Q_q=0,\pm1$ the corresponding charge of $(gg)$
in a gluon jet should be $Q_g=0$ \cite{Minkowski:2000qp}. Remarkably,
all the studies at LEP  by the experiments ALEPH, DELPHI 
and OPAL \cite{Schael:2006ns} 
have shown consistently, that the ``standard'' Monte Carlo JETSET without 
glueballs included is able to
reproduce the charge properties of quark jets but failed to describe the
charge distribution of leading clusters in gluon jets: 
there was an excess of neutral
charge $Q_g$. An example is shown in figure~\ref{fig:Q-Aleph} for the ALEPH Collaboration
with an excess at $Q_g=0$ of about 40\%.
Such an effect is just what is expected for glueballs, so we
take these observations as an indirect evidence for gluonic objects.
Alternatively, the MC program has modeled the gluon jet incorrectly.

In any case it would be desirable to clarify this issue at the LHC
\cite{Minkowski:2000qp,Ochs:2013gi}. Here the
higher jet energy allows for larger rapidity gaps and there 
is a high production rate
of gluon jets to be compared with quark jets.
Alternatively, one may compare the Feynman x-distribution of a resonance
with composition in doubt (example $f_0(1500))$ 
with a known ``test state'' (examples $\rho(770)$,
$f_2(1270)$, $\phi(1020)$). In a quark $u,d$ jet the leading 
$\rho$ would be ``harder''
than a glueball, whereas in a gluon jet the behaviour should be opposite.  

\section{Scalar meson spectroscopy with glueballs}
Most searches for glueballs have concentrated on the scalar sector
as the lightest glueball is a scalar according to QCD. A scalar glueball could mix into the
following states listed in \cite{beringer2012pdg}
\begin{equation}
f_0(500),\ f_0(980),\ f_0(1370),\ f_0(1500),\ f_0(1710)\ldots
\label{scalars}
\end{equation}
and there are various such mixing schemes. According to an early 
suggestion the heaviest three states in (\ref{scalars}) are mixed from
the two members of a $q\bar q$ nonet and a glueball centered around 1500
MeV \cite{Amsler:1995td}. Then there could be another nonet including the two 
$f_0$ states below
1 GeV together with $a_0(980)$ and the $K^*_0(900)/\kappa$ particle. 
A problem here is with
$f_0(1370)$ whose existence is in doubt as there is no evidence found in
the uniquely identified energy-independent phase shift solutions 
(for a recent review of the studied channels, see \cite{Ochs:2013gi}); also
the situation with $\kappa$ is difficult to clarify 
\cite{beringer2012pdg,Ochs:2013gi}. 
According to another suggestion the $q\bar q$ nonet includes $f_0(980)$
and $f_0(1500)$ while the broad $f_0(500)$ with ``Breit-Wigner mass'' 
$m_{BW}\sim 1000$ MeV, where the phase shifts pass $90^\circ$,
is related to the glueball \cite{Minkowski:1998mf}, related works in 
\cite{Narison:1988ts,Klempt:1995ku,Anisovich:2002ij}.

In \cite{Ochs:2013gi} an attempt has been made to determine the gluonic
component as given by the gluonic mixing angle $\phi_G$ 
and the flavour mixing angle $\phi_{sc}$ of scalars from
experimental data, for example
\begin{equation}
|f_0(1500)\ran=\cos \phi_G |q\bar q\ran + \sin \phi_G |gg\ran; \quad
 |q\bar q\ran =\cos \phi_{sc}|n\bar n\ran - \sin \phi_{sc} |s\bar s\ran.
\label{f01500mix}
\end{equation}
The following results have been obtained:

\begin{itemize}

\item $f_0(980)$: the scalar mixing angle (defined as in \cite{Ochs:2013gi}) is found as
$\phi_{sc}=(30\pm3)^\circ$ (similar to $\eta'$ with
$\phi_{ps}\sim 42^\circ$) from 
$\pi\pi\to \pi\pi/K\bar
K;\ J/\psi\to (\omega/\phi)\pi\pi, D_s\to 3\pi,\ \gamma\gamma\to
f_0(980)/a_0(980)$; there is no definitive 
evidence for a gluonic contribution 
($\cos^2\phi_G\approx 0.75\pm 0.25$).

\item $f_0(500)$: couplings $r_{K\pi}\equiv g_{K\bar K}^2/g_{\pi\pi}^2=0.3-0.6$ 
and $r_{\eta\pi}\equiv g_{\eta\eta}^2/g_{\pi\pi}^2\sim 0.2$ based on model fits
to $\pi\pi\to \pi\pi/K \bar K/4\pi/\eta\eta$ amplitudes
\cite{Kaminski:2009qg,Bugg:2006sr} suggest  
this state to be neither pure glue ($r_{K\pi} 
=4/3$) nor pure nonstrange $qq\bar q\bar q$ ($r_{K\pi}=0$).

\item $f_0(1500)$: the 2-body branching ratios in \cite{beringer2012pdg} 
do not fix $\phi_{sc}$; given
$\phi_{sc}$ the gluon component $r_G=r^0_G\tan \phi_G$ is fixed where
$r_G^0=G_0/g_0$ refers to decay couplings $G_0,\ g_0$ of 
($gg)$ and ($q\bar q$).
\end{itemize}
In a minimal mixing scheme  \cite{Ochs:2013gi} 
one assumes $f_0(980)$ without glue and
with $\phi_{sc}=30^\circ$, the glueball is mixed into $f_0(1500)$ and
$f_0(500)$ with 
\begin{equation}
|f_0(500)\ran=\sin \phi_G |q\bar q\ran - \cos \phi_G |gg\ran; \quad
 |q\bar q\ran =\cos \phi_{sc}|n\bar n\ran - \sin \phi_{sc}|s\bar s\ran.
\label{f0500mix}
\end{equation}
With the above experimental results for  $f_0(500)$ and $f_0(1500)$ we obtain 
\begin{equation}
\phi_G\sim 50^\circ, \qquad r_G^0\approx 0.5
\label{phiGmix}
\end{equation}
i.e. a strongly mixed system. Note that for $\phi_G=0$ we recover the model
\cite{Minkowski:1998mf}, while the composition of $f_0(1500)$ is similar to
the model \cite{Amsler:1995td} but with a different $q\bar q$ multiplet. 
An interesting aspect of this mixing scheme is the compatibility with
certain
results obtained from lattice QCD and from QCD sum rules (see section 1). 

Further tests on the flavour composition are possible from the decays 
$B\to J/\psi f_0$ and $B_s\to J/\psi f_0$ which allow for an isolation of 
the $d\bar d$ and the $s\bar s$ components of $f_0$ respectively. 
From $B_s$ decays using equations (\ref{f01500mix}, \ref{f0500mix}) 
one can obtain the ratio of the $s\bar s$ components as 
\begin{equation}
r_{f_0'/f_0}^s=\frac{p^3 B( \bar B^0_s\to J/\psi f_0(1500))}{(p')^3 
B(\bar B^0_s\to J/\psi f_0(980))},\quad 
   r_{f_0'/f_0}^s=\cos^2\phi_G^{f_0'}\tan^2\phi_{sc} 
\label{rf0}
\end{equation}
with $p(p')$ the respective momenta in the $B_s$ cms. 
The LHCb results for $f_0\to \pi\pi$ \cite{LHCb:2012ae} 
yield two solutions of the phase shift analysis, from the recent data
on the $f_0\to K^+K^-$ decay mode \cite{Aaij:2013orb}
we conclude that the ''alternate solution" is preferred with
$r_{f_0'/f_0}^s=0.15\pm0.05$; this supersedes our choice in
\cite{Ochs:2013gi}. Then we derive from (\ref{rf0})
using $\phi_{sc}=30^\circ$ the gluonic mixing
angle $\phi_G^{f_0'}=48^\circ\pm10^\circ$ in good agreement with
our result above in (\ref{phiGmix}). 

The flavour mixing angle $\phi_{sc}$ for a particular $f_0$ meson can be
derived from the ratio of the $J/\psi f_0$ decays of $B$ and $B_s$
\cite{Ochs:2013gi}. First indications of a $f_0(1500)$ signal in $B_d$
decays become visible \cite{Aaij:2013zpt}. 
A measurement of
these ratios for the light scalars $f_0$ would greatly advance our
understanding of this sector. Note that $f_0(500)$ in our view is to be
represented in a phase shift analysis as broad object centered at
$m_{BW}\sim 1000$ MeV \cite{Minkowski:1998mf,Ochs:2013gi}.  

\section{Nonets and glueballs from symmetry relations}
Another area of promising research are the decay rates of
charmonia (or bottomonia). In the decay $\chi_c\to gg$ the
pairs of hadrons from the same $q\bar q$ multiplet form a flavour singlet
state.
For pairs of pseudoscalars one expects 
for the phase space corrected decay rates 
\begin{equation}
 \gamma^2_{ij}(\chi_c\to\pi\pi:K\bar K:\eta\eta:\eta\eta':\eta'\eta')
\quad  = \quad3:4:1:0:1 \label{statweights}
\label{su3symmetry}
\end{equation}
reflecting the $SU(3)_{fl}$ symmetry. These relations 
work rather well for decays of $\chi_{c0}(0^{++})$ and 
$\chi_{c2}(0^{++})$ within 10-20\% \cite{Ochs:2013gi}. Correspondingly,
one may test this symmetry for decays into members of other multiplets, 
in particular of the scalar multiplet. The relations in analogy to   
(\ref{su3symmetry}) are expected 
 to fail if the wrong members of the multiplet are chosen, furthermore,
deviations are \linebreak
expected for the isoscalar members in the presence of glueballs.
The PDG results \cite{beringer2012pdg} (based on BES-II
\cite{Ablikim:2005kp}) yield the ratio of
decay rates after correction for phase space $q$ and charge weight $c$
\begin{equation}
r_{f_0/K^*}^{\chi_{c0}}\equiv \frac{B(\chi_{c0}(3415)\to
f_0(980)f_0(980))/(cq)}
{ B(\chi_{c0}(3415)\to K_0^*(1430)\bar K_0^*(1430))/(c'q')}, 
       \quad r_{f_0/K^*}^{\chi_{c0}}=1.64\pm0.73
\end{equation}
which is consistent with symmetry ($r_{f_0/K^*}^{\chi_{c0}}=1$) and so with
$f_0(980)$ and $K^*_0(1430)$ being in the same $q\bar q$ nonet
\cite{Klempt:1995ku,Minkowski:1998mf}. Higher accuracy and
measurements of relations to other scalars like $a_0(980), f_0(1500)$ and
also to $f_0(500), K^*_0(900)/\kappa$ could be decisive. The results also
prefer $f_0(980)$ to be a ($q\bar q$) rather than a ($qq\bar q\bar q$) 
state according to  \cite{Ochs:2013gi} from a comparison with $p\bar p$ data.
%
%
\section{Summary}
QCD predicts the existence of glueballs; mass and mixing properties of
the scalar glueball need further clarification. Experimental hints for
glueballs come from the excess of neutral clusters in gluon jets.
The intrinsic quark-gluon structure of mesons is accessible experimentally;
the role of $B,\ B_s$ data is stressed. Tests of symmetry relations for 
$\chi_c$ decays help identifying $q\bar q$ nonets.


\begin{thebibliography}{99}
\bibitem{Ochs:2013gi} 
  W.~Ochs,
 ``The Status of Glueballs,''
  J.\ Phys.\ G {\bf 40} (2013) 043001

\bibitem{Fritzsch:1975tx}
  H.~Fritzsch and P.~Minkowski,
  Nuovo Cim.\ A {\bf 30} (1975) 393

\bibitem{Richards:2010ck}
  C.~M.~Richards {\it et al.}  [UKQCD Collaboration],
  Phys.\ Rev.\ D {\bf 82} (2010) 034501

\bibitem{Hart:2006ps}
  A.~Hart {\it et al.}  [UKQCD Collaboration],
  Phys.\ Rev.\ D {\bf 74} (2006) 114504

\bibitem{Narison:1988ts}
  S.~Narison and G.~Veneziano,
  Int.\ J.\ Mod.\ Phys.\ A {\bf 4} (1989) 2751
\bibitem{Harnett:2008cw}
  D.~Harnett, R.~T.~Kleiv, K.~Moats and T.~G.~Steele,
  Nucl.\ Phys.\ A {\bf 850} (2011) 110


\bibitem{Roy:1978pz}
  P.~Roy and T.~F.~Walsh,
  Phys.\ Lett.\ B {\bf 78} (1978) 62
\bibitem{Minkowski:2000qp}
  P.~Minkowski and W.~Ochs,
  Phys.\ Lett.\ B {\bf 485} (2000) 139;
%
 XIV Int. Conf. on Hadron Specrtroscopy (Hadron 2011), 13-17 June 2011,
 Munich, Germany),  arXiv:1108.0589  [hep-ph]

\bibitem{Schael:2006ns}
  S.~Schael {\it et al.}  [ALEPH Coll.],
  Eur.\ Phys.\ J.\ C {\bf 48} (2006) 685;
  J.~Abdallah {\it et al.}  [DELPHI Coll.],
  Phys.\ Lett.\ B {\bf 643} (2006) 147;
  G.~Abbiendi {\it et al.}  [OPAL Coll.],
  Eur.\ Phys.\ J.\ C {\bf 35} (2004) 293



\bibitem{beringer2012pdg}
J.~Beringer et al. (Particle Data Group),
Phys.\ Rev.\ D {\bf 86} (2012) 010001 

\bibitem{Amsler:1995td}
  C.~Amsler and F.~E.~Close,
  Phys.\ Rev.\ D {\bf 53} (1996) 295

\bibitem{Minkowski:1998mf}
  P.~Minkowski and W.~Ochs,
  Eur.\ Phys.\ J.\ C {\bf 9} (1999) 283

\bibitem{Klempt:1995ku}
  E.~Klempt et al.  
  Phys.\ Lett.\ B {\bf 361} (1995) 160;  
%
  V.~Dmitrasinovic,
  Phys.\ Rev.\ C {\bf 53} (1996) 1383

\bibitem{Anisovich:2002ij}
  V.~V.~Anisovich and A.~V.~Sarantsev,
  Eur.\ Phys.\ J.\ A {\bf 16} (2003) 229

\bibitem{Kaminski:2009qg}
  R.~Kaminski, G.~Mennessier and S.~Narison,
  Phys.\ Lett.\ B {\bf 680} (2009) 148

\bibitem{Bugg:2006sr}
  D.~V.~Bugg,
  Eur.\ Phys.\ J.\ C {\bf 47} (2006) 45

\bibitem{LHCb:2012ae}
  R.~Aaij {\it et al.}  [LHCb Collaboration],
Phys.\ Rev.\ D {\bf 86} (2012) 052006

\bibitem{Aaij:2013orb}
  R.~Aaij {\it et al.}  [LHCb Collaboration],
  Phys.\ Rev.\ D {\bf 87} (2013) 072004

\bibitem{Aaij:2013zpt}
  R.~Aaij {\it et al.}  [LHCb Collaboration],
  Phys.\ Rev.\ D {\bf 87} (2013),  052001 and
%
  arXiv:1310.2145 

\bibitem{Ablikim:2005kp} 
  M.~Ablikim {\it et al.}  [BES Collaboration],
  Phys.\ Rev.\ D {\bf 72} (2005) 092002;
%
\ D {\bf 70} (2004) 092002



\end{thebibliography}
\end{document}